\documentclass[slac_one]{revtex4}
\usepackage{graphicx}
\usepackage{fancyhdr}
\newcommand{\be}{\begin{equation}}
\newcommand{\ee}{\end{equation}}
\newcommand{\bea}{\begin{eqnarray}}
\newcommand{\eea}{\end{eqnarray}}

\newcommand{\MS}{\ensuremath{\overline{\text{MS}}}}

\begin{document}


\title {\centering{ The $\bar{B}\to X_s\gamma \gamma$ decay: \\
 NLL QCD contribution of the Electromagnetic
  Dipole operator $\mathcal{O}_{7}$    }}

%

%
    %
\author{ \bf Ahmet Kokulu }
\email[Electronic address:]{akokulu@itp.unibe.ch}
\affiliation{ \sl   Albert Einstein Center for Fundamental Physics,
 Institute for Theoretical Physics,
 Univ. of Bern, CH-3012 Bern, Switzerland   }    %
%

\begin{abstract}
  ~~We calculate the set of $O(\alpha_s)$ corrections to the double 
differential decay width
$d\Gamma_{77}/(ds_1 \, ds_2)$
for the process $\bar{B} \to X_s \gamma \gamma$ originating from
diagrams involving the electromagnetic dipole operator
${\cal O}_7$. The kinematical variables $s_1$ and $s_2$ are defined as
$s_i=(p_b - q_i)^2/m_b^2$, where $p_b$, $q_1$, $q_2$ are the momenta
of $b$-quark and two photons. While the (renormalized) virtual
corrections are worked out exactly for a certain range of $s_1$ and $s_2$, 
we retain in the gluon bremsstrahlung process only the leading power w.r.t. the (normalized)
hadronic mass $s_3=(p_b-q_1-q_2)^2/m_b^2$ in the underlying triple
differential decay width $d\Gamma_{77}/(ds_1 ds_2 ds_3)$. The double
differential decay width, based on this approximation, is free of
infrared- and collinear singularities when combining virtual- and
bremsstrahlung corrections. The corresponding results are obtained 
analytically. When retaining all powers in $s_3$, the sum of virtual-
and bremstrahlung corrections contains uncanceled $1/\epsilon$
singularities (which are due to collinear {\it photon} emission from the
$s$-quark) and other concepts, which go beyond
perturbation theory, like parton fragmentation functions of a quark or
a gluon into a photon, are needed which is beyond the scope of our paper.
\end{abstract}

\maketitle

\thispagestyle{fancy}


\section{Introduction}
Inclusive rare $B$-meson decays are known to be a unique source of indirect information about 
physics at scales of several hundred GeV. In the Standard Model (SM) all these processes 
proceed through loop diagrams and thus are relatively suppressed. In the extensions 
of the SM the contributions stemming from the diagrams with ``new'' 
particles in the loops can be comparable or even larger than the contribution from 
the SM. Thus getting experimental information on rare decays puts strong 
constraints on the extensions of the SM or can even lead to a  
disagreement with the SM predictions, providing evidence for some ``new physics''. 

To make a rigorous comparison between experiment and theory, precise
SM calculations for the (differential) decay rates are mandatory. While the
branching ratios for $\bar{B} \to X_s \gamma$ \cite{Misiak:2006zs}
and $\bar{B} \to X_s \ell^+
\ell^-$ are known today even to
next-to-next-to-leading logarithmic (NNLL) precision (for reviews, see
\cite{Hurth:2010tk,Buras:2011we}),
other branching ratios, like the one for $\bar{B} \to X_s \gamma
\gamma$ discussed in these proceedings, has been calculated before to leading logarithmic
(LL) precision in the SM by several groups \cite{Simma:1990nr,Reina:1996up,Reina:1997my,Cao:2001uj} and  only recently a first step towards next-to-leading-logarithmic (NLL) precision was presented by us in \cite{Asatrian:2011ta}.
In contrast to $\bar{B} \to X_s
\gamma$, the current-current operator ${\cal O}_2$ has a non-vanishing matrix
element for $b \to s \gamma \gamma$ at order $\alpha_s^0$ precision, 
leading to an interesting interference pattern with the contributions associated
with the electromagnetic dipole operator ${\cal O}_7$ already at LL
precision. As a consequence, potential new physics should be clearly visible
not only in the total branching ratio, but also in the 
differential distributions.
 
As the process $\bar{B} \to X_s \gamma \gamma$ is expected to be measured at
the planned Super $B$-factories in Japan and Italy, it is necessary
to calculate the differential distributions to NLL precision in the
SM, in order to
fully exploit its potential concerning new physics. 
The starting point of our calculation is the effective Hamiltonian,
obtained by integrating out the heavy particles in the SM, leading to
\be
 {\cal H}_{eff} = - \frac{4 G_F}{\sqrt{2}} \,V_{ts}^\star V_{tb} 
   \sum_{i=1}^8 C_i(\mu) {\cal O}_i(\mu)  \, ,
\label{Heff}
\ee
where we use the operator basis introduced in \cite{Chetyrkin:1996vx}:
\be
\begin{array}{llll}
{\cal O}_1 \,= &\!
 (\bar{s}_L \gamma_\mu T^a c_L)\, 
 (\bar{c}_L \gamma^\mu T_a b_L)\,,
               &  
{\cal O}_2 \,= &\!
 (\bar{s}_L \gamma_\mu c_L)\, 
 (\bar{c}_L \gamma^\mu b_L)\,,   \\[1.002ex]
{\cal O}_3 \,= &\!
 (\bar{s}_L \gamma_\mu b_L) 
 \sum_q
 (\bar{q} \gamma^\mu q)\,, 
               & 
{\cal O}_4 \,= &\!
 (\bar{s}_L \gamma_\mu T^a b_L) 
 \sum_q
 (\bar{q} \gamma^\mu T_a q)\,,  \\[1.002ex]
{\cal O}_5 \,= &\!
 (\bar{s}_L \gamma_\mu \gamma_\nu \gamma_\rho b_L) 
 \sum_q
 (\bar{q} \gamma^\mu \gamma^\nu \gamma^\rho q)\,, 
               & 
              {\cal O}_6 \,= &\!
 (\bar{s}_L \gamma_\mu \gamma_\nu \gamma_\rho T^a b_L) 
 \sum_q
 (\bar{q} \gamma^\mu \gamma^\nu \gamma^\rho T_a q)\,,  \\[1.002ex]
{\cal O}_7 \,= &\!
  \frac{e}{16\pi^2} \,\bar{m}_b(\mu) \,
 (\bar{s}_L \sigma^{\mu\nu} b_R) \, F_{\mu\nu}\,, 
               &  
{\cal O}_8 \,= &\!
  \frac{g_s}{16\pi^2} \,\bar{m}_b(\mu) \,
 (\bar{s}_L \sigma^{\mu\nu} T^a b_R)
     \, G^a_{\mu\nu}\, .   
\end{array} 
\label{opbasis}
\ee

The symbols $T^a$ ($a=1,8$) denote the $SU(3)$ color generators; 
$g_s$ and $e$, the strong and electromagnetic coupling constants.
In eq.~(\ref{opbasis}), 
$\bar{m}_b(\mu)$ is the running $b$-quark mass 
in the $\MS$-scheme at the renormalization scale $\mu$.
As we are not interested in CP-violation effects in the present paper, we 
made use of the hierarchy
 $V_{ub} V_{us}^* \ll V_{tb} V_{ts}^* $ when writing
eq. (\ref{Heff}). Furthermore, we also put $m_s=0$.  

While the Wilson coefficients $C_i(\mu)$ appearing in eq. (\ref{Heff})
are known to sufficient precision at the low scale $\mu \sim m_b$
since a long time (see e.g. the reviews \cite{Hurth:2010tk,Buras:2011we}
and references therein), the matrix elements 
$\langle s \gamma \gamma|{\cal  O}_i|b\rangle$ and 
$\langle s \gamma \gamma \, g|{\cal  O}_i|b\rangle$, 
which in a NLL calculation are needed to order
$g_s^2$ and $g_s$, respectively, are not known yet. To calculate the
$({\cal O}_i,{\cal O}_j)$-interference contributions to the
differential distributions at order
$\alpha_s$ is in many respects of similar complexity as the
calculation of the photon energy spectrum in $\bar{B} \to X_s \gamma$ 
at order $\alpha_s^2$
needed for the NNLL computation. As a first step in this NLL enterprise, we
derived in our paper \cite{Asatrian:2011ta}, the $O(\alpha_s)$
corrections 
to the $({\cal O}_7,{\cal O}_7)$-interference contribution to the double 
differential decay width $d\Gamma/(ds_1 ds_2)$ at the partonic level. 
The variables $s_1$
and $s_2$ are defined as $s_i=(p_b-q_i)^2/m_b^2$, where $p_b$
and $q_i$ denote the four-momenta of the $b$-quark and the two
photons, 
respectively.

At order $\alpha_s$
there are contributions to $d\Gamma_{77}/(ds_1 ds_2)$ with three
particles 
($s$-quark and two photons)
and four particles ($s$-quark, two photons and a gluon) in the final state.
These contributions correspond to specific cuts of the $b$-quarks
self-energy at order $\alpha^2 \times \alpha_s$, involving twice the
operator ${\cal O}_7$. As there are additional cuts, which contain for
example only one photon, our observable cannot be obtained using the
optical theorem, i.e., by taking the absorptive part of the $b$-quark
self-energy at three-loop. We therefore calculate the mentioned 
contributions with three and four particles in the final state individually.

We work out the QCD corrections
to the double differential decay width 
in the kinematical range 
\[
0 < s_1 < 1 \quad ; \quad 0 < s_2 < 1-s_1 \, .
\]

Concerning the virtual corrections, all singularities (after
ultra-violet renormalization) are due to  {\bf soft gluon} exchanges
and/or  {\bf collinear gluon} exchanges involving the $s$-quark. Concerning the
bremsstrahlung corrections (restricted to the same range of $s_1$ and
$s_2$), there are in addition kinematical situations where {\bf collinear photons}
are emitted from the $s$-quark. The corresponding singularities are not
canceled when combined with the virtual corrections. We found, however, that there are no
singularities associated with collinear photon emission in the double
differential decay width when only retaining
the leading power w.r.t to the (normalized) hadronic mass
$s_3=(p_b - q_1 - q_2)^2/m_b^2$ in the underlying triple differential distribution
$d\Gamma_{77}/(ds_1 ds_2 ds_3)$. Our results of our paper are
obtained within this ``approximation''. When going beyond, other
concepts which go beyond perturbation theory, like parton
fragmentation functions of a quark or a gluon into a photon, are
needed \cite{Kapustin:1995fk}.

\section{Leading Order and Final results for the decay width }\label{sec:combination}
In $d=4$ dimensions, the leading-order spectrum (in our restricted
phase-space) is given by
\begin{eqnarray}
&&\frac{d\Gamma_{77}^{(0)}}{ds_1 \, ds_2} = \frac{\alpha^2 \, \bar{m}_b^2(\mu) \, m_b^3 \, |C_{7,eff}(\mu)|^2 \, G_F^2 \,
  |V_{tb} V_{ts}^*|^2 \,  Q_d^2}{1024 \, \pi^5} \, 
  \frac{(1-s_1-s_2)}{(1-s_1)^2 s_1 (1-s_2)^2 s_2} \, r_0 \, .
\label{treezero}
\end{eqnarray}
where
\begin{eqnarray}
\nonumber r_0&=&-48 s_2^3 s_1^3+96 s_2^2 s_1^3-56 s_2 s_1^3+8
   s_1^3+96 s_2^3 s_1^2-192 s_2^2 s_1^2+112
   s_2 s_1^2-56 s_2^3 s_1+
   \\&&\nonumber 112 s_2^2 s_1-96 s_2
   s_1+8 s_1+8 s_2^3+8 s_2
    \end{eqnarray}

The complete order $\alpha_s$ correction to the
double differential decay width
$d\Gamma_{77}/(ds_1 \, ds_2)$ is obtained by
adding the renormalized virtual corrections and the bremsstrahlung corrections. Explicitly we obtain
\begin{eqnarray}
 \frac{d\Gamma_{77}^{(1)}}{ds_1 \, ds_2} = 
\frac{\alpha^2 \, \bar{m}_b^2(\mu) \, m_{b}^3 \, |C_{7,eff}(\mu)|^2 \, G_F^2 \,
  |V_{tb} V_{ts}^*|^2 \,  Q_d^2}{1024 \, \pi^5} \times
 \frac{\alpha_s}{4\pi} \, C_F  \, \left[
\frac{-4 \, r_0 \, (1-s_1-s_2)}{(1-s_1)^2 \, s_1 \, (1-s_2)^2 \, s_2} \, \log \frac{\mu}{m_b} +f
\right]  \, ,
\label{total}
\end{eqnarray}
where $f$ can be found explicitly in \cite{Asatrian:2011ta}. 

The order $\alpha_s$ correction $d\Gamma_{77}^{(1)}/(ds_1 ds_2)$ in
Eq. (\ref{total}) to the double differential decay width
for $b \to X_s \gamma \gamma$ was the
main result of our paper \cite{Asatrian:2011ta}.

\section{Some numerical illustrations}\label{sec:numerics}
\begin{figure}[h]
\centering{
\includegraphics[width=0.34\textwidth]{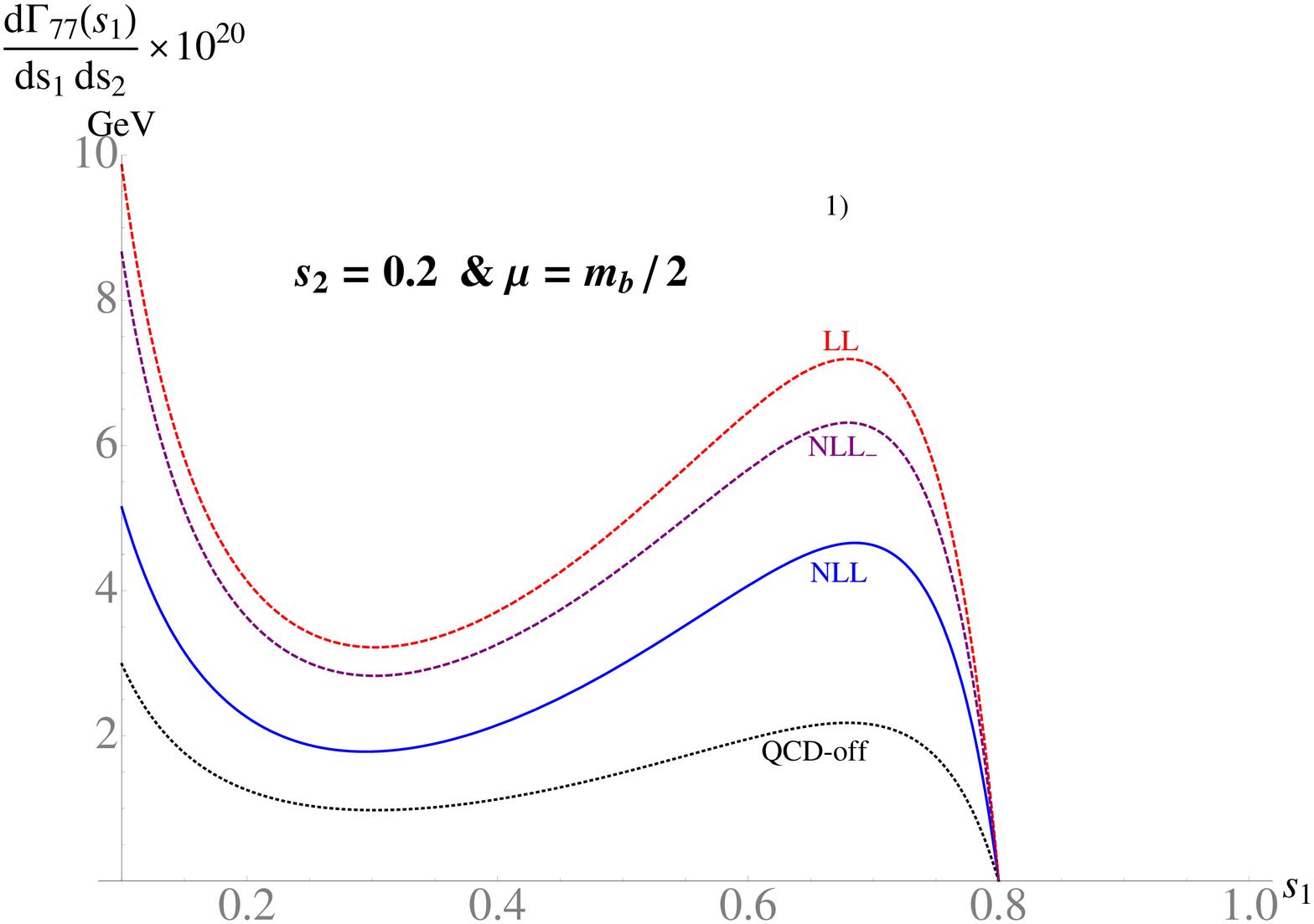}
\hspace{0.9cm}
\includegraphics[width=0.34\textwidth]{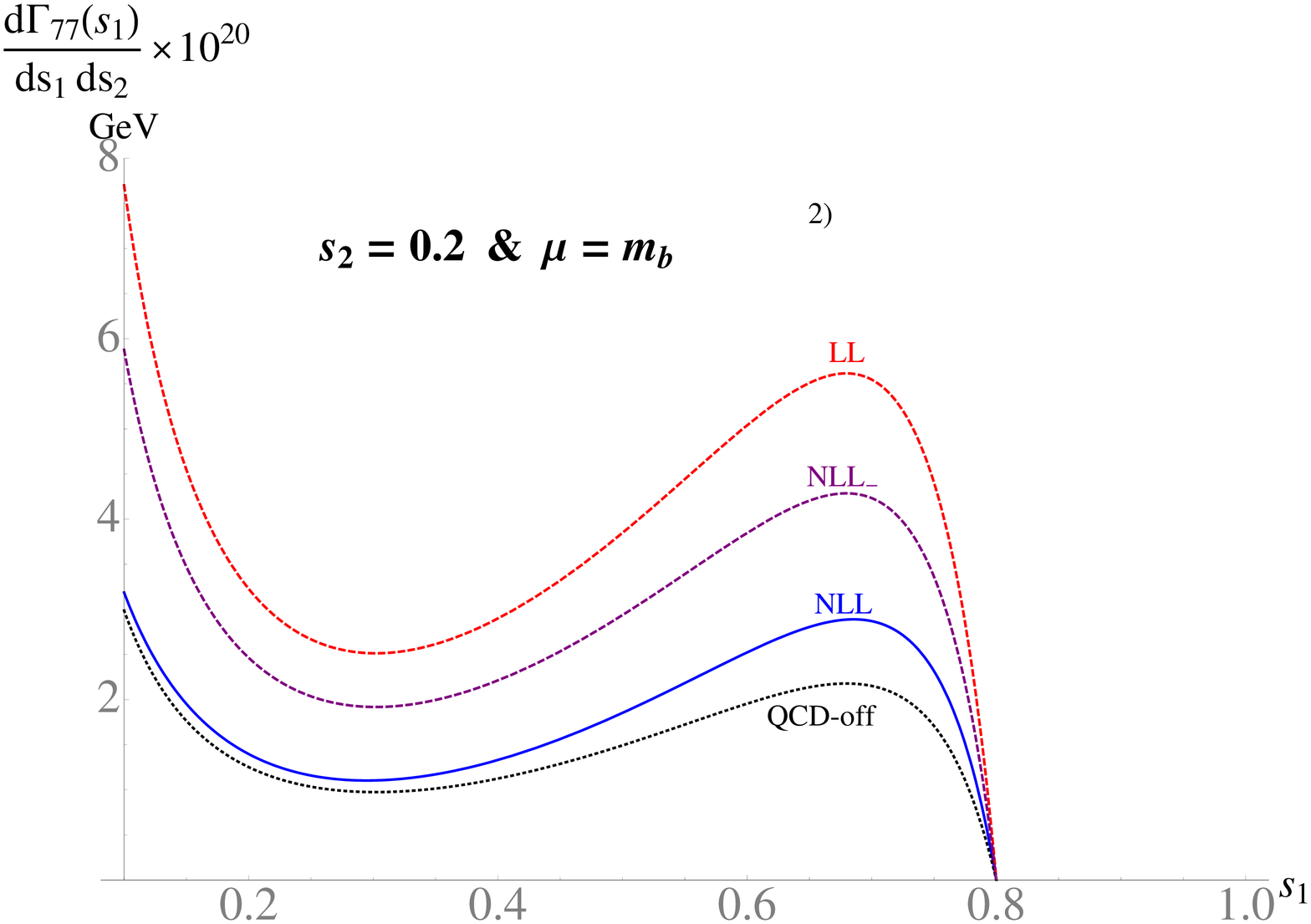}    \\
\vspace{0.2cm}
\includegraphics[width=0.34\textwidth]{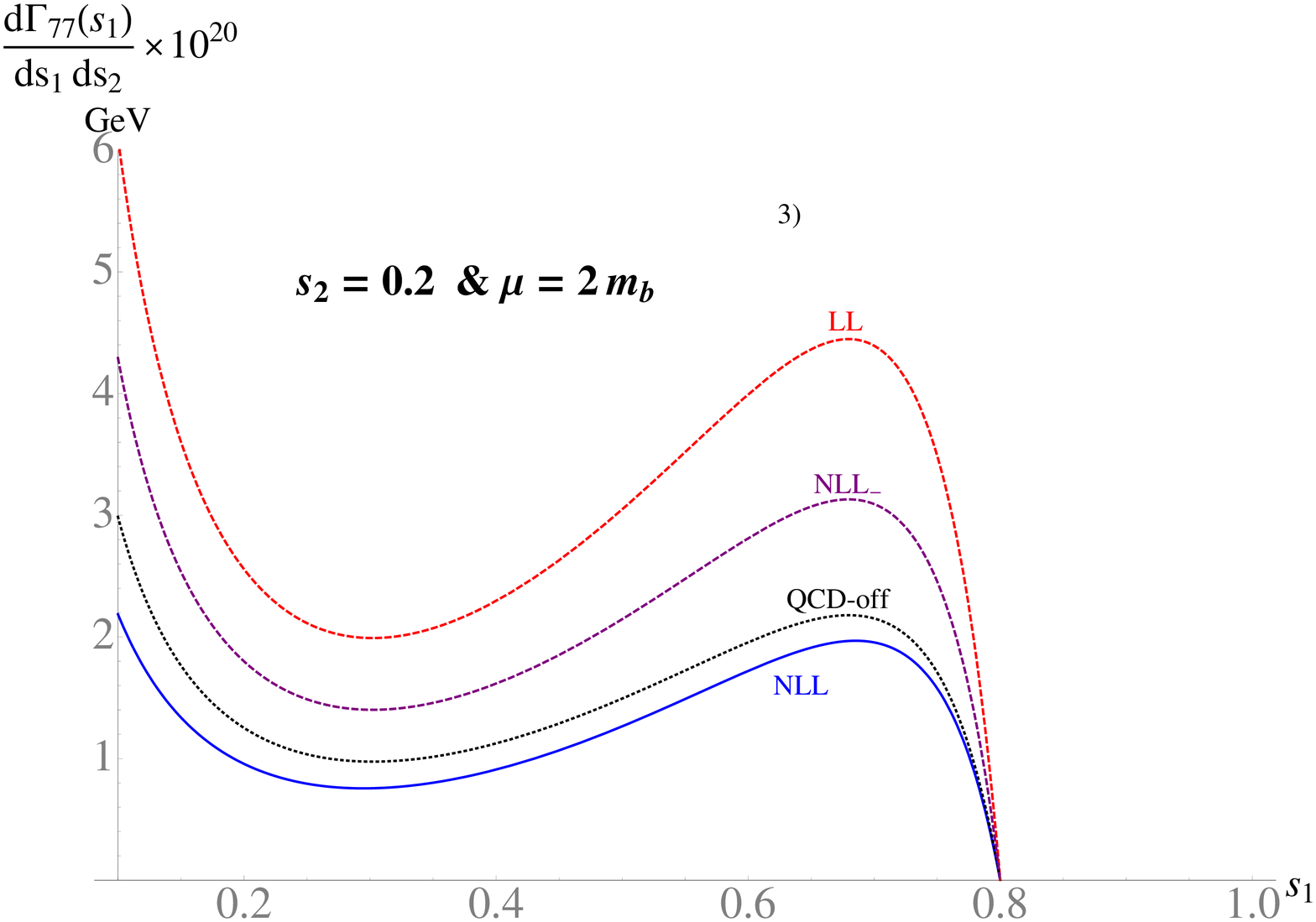}~~~~~~~~~~~~~~~~~~~~~~~~~~~~~~~~
\includegraphics[width=0.17\textwidth]{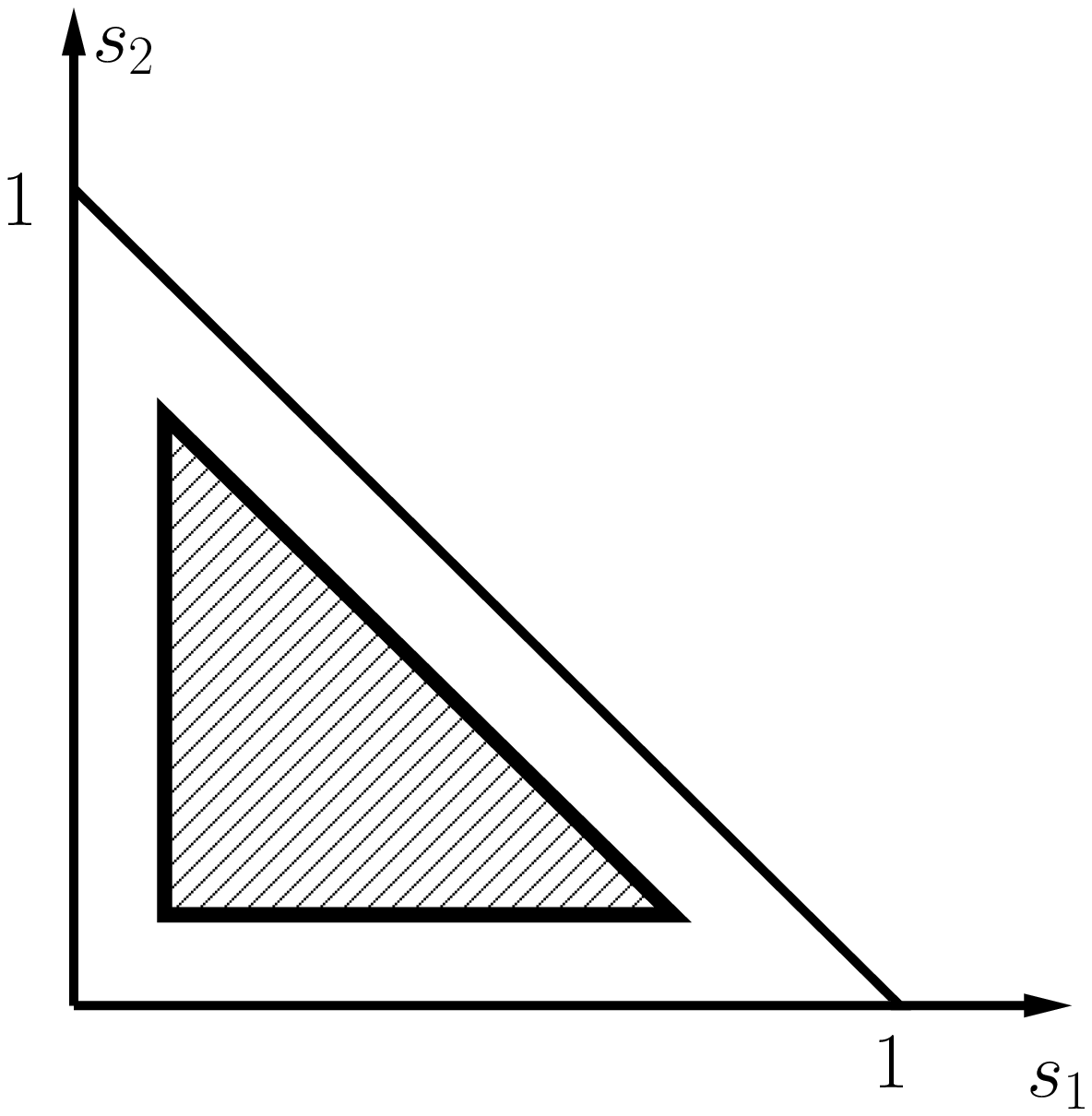}   }  
\caption{\footnotesize{{\bf Frames 1)-3)}: Double differential decay width $d\Gamma_{77}/(ds_1 ds_2)$ 
as a function of $s_1$ for $s_{2}$ fixed at $s_2=0.2$. 
The dotted(black), the short-dashed(red) and the solid line(blue) shows the result 
when neglecting QCD-effects, the LL and the NLL result,
respectively. The long-dashed line(purple) represents the (partial) NLL result
in which the virtual- and bremsstrahlung corrections worked out in our
paper \cite{Asatrian:2011ta} are switched off. In the frames 1), 2) and 3) the renormalization scale is chosen to
  be $\mu=m_b/2$, $\mu=m_b$ and $\mu=2 \, m_b$, respectively. {\bf Down right:} The relevant phase-space region for $s_{1}$ and $s_{2}$.}}
\label{fig:results}
\end{figure}

In our procedure the NLL corrections have three sources: 
(a) $\alpha_s$ corrections to the Wilson coefficient $C_{7,eff}(\mu)$,
(b) expressing $\bar{m}_b(\mu)$ in terms of the pole mass $m_b$ and
(c) virtual- and real- order $\alpha_s$ corrections to the matrix
elements. To illustrate the effect of source (c), which is worked out
for the first time in our paper \cite{Asatrian:2011ta}, we show in Fig. \ref{fig:results}
(by the long-dashed line) the (partial) NLL result in which 
source (c) is switched off. 
We conclude that the effect (c) is roughly of equal
importance as the combined effects of (a) and (b). 

For completeness we
show in this figure (by the dotted line) also the result when QCD
is completely switched off, which amounts to put $\mu=m_W$ in the LL
result. 

From Fig. \ref{fig:results} we see that the NLL results are
substantially smaller (typically by $50\%$ or slightly more) than
those at LL precision, which is also the case when choosing other
values for $s_2$. 

In the numerical discussion above, we have systematically converted the running
$b$-quark mass $\bar{m}_b(\mu)$ in terms of the pole mass $m_b$.
As perturbative expansions often behave better when expressed in terms
of the running  mass, we also
studied the results obtained when systematically converting $m_b$ 
in terms of  $\bar{m}_b(\mu)$. After doing also this version, 
we observe the following:
Generally speaking, NLL corrections are not small
for both cases, when taking
into account the full range of $\mu$, i.e., $m_b/2<\mu<2 \, m_b$.
More precisely, in   the $\overline{\mbox{MS}}$ version they are  large
for $\mu=m_b/2$ and smaller for larger values of $\mu$, while in the pole
mass version they are large for all values of $\mu$.

We stress that the numerically important contributions
involving the operator $O_2$ are not discussed in our paper. Therefore, the issue concerning the
reduction of the $\mu$ dependence at NLL precision cannot be addressed
at this level. Finally, the relevant input parameters that we used in our analysis together with the values of the Wilson coefficient $C_{7}$ and the strong coupling $\alpha_{s}$ at different values of the scale $\mu$ are listed in Table I.


\begin{table}[htdp]
  \label{tab1:wilson}
\footnotesize{\begin{minipage}{2in}
\centering 
 \tiny
  \begin{tabular}{@{}|c|c|c|c@{}|}
 \hline
 Parameter & Value \\   \hline \hline

$\rm m_{b}({pole})$& $4.8$~GeV   \\ \hline

$\rm m_{t}({pole})$&$175$~GeV    \\ \hline

$\rm M_{W}$&$80.4$~GeV    \\ \hline

$\rm M_{Z}$&$91.19$~GeV   \\ \hline

$\rm G_{F}$&$1.16637\times10^{-5}$~\text{GeV}$^{-2}$  \\ \hline

$\rm V_{tb} V_{ts}^* $&$0.04$    \\  \hline

$\rm {\alpha}^{-1}$&$137$    \\ \hline

$\rm {\alpha_{s}(M_{Z})}$&$0.119$      \\ \hline \hline
\end{tabular}
 \end{minipage}}~~~~~~~~~~~~~
\footnotesize{\begin{minipage}{2in}
\tiny
   \begin{tabular}{@{}|c|c|c|c@{}|}
\hline
  &  $\alpha_s(\mu)$ & $C_{7,eff}^{0}(\mu)$ & $C_{7,eff}^{1}(\mu)$
\\   \hline \hline

$\mu= \rm M_W$  & $0.1213$ & $-0.1957$ & $-2.3835$ \\ \hline 

$\mu=2 \, m_{b}$  & $0.1818$ & $-0.2796$ & $-0.1788$ \\ \hline 

$\mu=m_{b}$  & $0.2175$ & $-0.3142$ &  $0.4728$   \\ \hline

$\mu=m_{b}/2$  & $0.2714$ & $-0.3556$ & $1.0794$    \\ \hline \hline
\end{tabular}
    \end{minipage}}
\caption{ \footnotesize{  {\bf Left:}  Relevant input parameters . {\bf Right:}  $\alpha_s(\mu)$ and the Wilson coefficient $C_{7,eff}(\mu)$ at  different values of the scale $\mu$.}}
\end{table}   
%

\section{Concluding remarks}\label{sec:summary}
In the present work we calculated the set of the $O(\alpha_s)$
corrections to the decay process $\bar{B} \to X_s \gamma \gamma$
originating from diagrams involving ${\cal O}_7$.
To perform this calculation, it is necessary to work out diagrams
with three particles ($s$-quark and two photons) and four
particles ($s$-quark, two photons and a gluon) in the final state.
From the technical point of view, the calculation was made possible
by the use of the Laporta Algorithm  to identify the
needed master integrals and by applying the differential equation method to
solve the master integrals.
When calculating the bremsstrahlung corrections, we take into account
only terms proportional to the leading power of the hadronic mass.
We find that the infrared and collinear singularities cancel when
combining the above mentioned approximated version of bremsstrahlung
corrections with the virtual corrections.
The numerical impact of the NLL corrections is large: for
$d\Gamma_{77}/(ds_1 \, ds_2)$ the NLL result is approximately
50\% smaller than the LL prediction.
\begin{acknowledgments}
I wish to thank to the organizers of the conference FPCP 2012 for their efforts to make it very pleasant.
This work is partially supported by the Swiss National Foundation and by
the Helmholz Association through 
funds provided to the virtual institute ``Spin and strong QCD''
(VH-VI-231).
\end{acknowledgments}

\bibliography{kokulu_FPCP12}


\end{document}